\newcommand{\CLASSINPUTtoptextmargin}{2.4cm}
\newcommand{\CLASSINPUTbottomtextmargin}{4.6cm}
\documentclass[conference]{IEEEtran}
\IEEEoverridecommandlockouts
\usepackage{cite}
\usepackage{amsmath,amssymb,amsfonts}
\usepackage{algorithmic}
\usepackage{graphicx}
\usepackage{textcomp}
\usepackage{xcolor}
\usepackage{tabularx}
\usepackage{booktabs}
\usepackage{graphicx}
\usepackage{subcaption}
\usepackage{multirow}
\usepackage{tabularray}
\usepackage{color}
\usepackage{pdfcomment}

\def\BibTeX{{\rm B\kern-.05em{\sc i\kern-.025em b}\kern-.08em
    T\kern-.1667em\lower.7ex\hbox{E}\kern-.125emX}}

\begin{document}


\title{
NMformer: A Transformer for Noisy Modulation Classification in Wireless Communication\\

}

    

\author{\IEEEauthorblockN{1\textsuperscript{st} Atik Faysal}
\IEEEauthorblockA{\textit{Electrical and Computer Engineering} \\
\textit{Rowan University}\\
Glassboro, NJ \\
faysal24@rowan.edu}
\and
\IEEEauthorblockN{2\textsuperscript{nd} Mohammad Rostami}
\IEEEauthorblockA{\textit{Electrical and Computer Engineering} \\
\textit{Rowan University}\\
Glassboro, NJ  \\
rostami23@students.rowan.edu}
\and
\IEEEauthorblockN{3\textsuperscript{rd} Reihaneh Gh. Roshan}
\IEEEauthorblockA{\textit{Computer Science} \\
\textit{Stevens Institute of Technology}\\
Hoboken, NJ \\
rghasemi@stevens.edu}
\and
\IEEEauthorblockN{4\textsuperscript{th} Huaxia Wang}
\IEEEauthorblockA{\textit{Electrical and Computer Engineering} \\
\textit{Rowan University}\\
Glassboro, NJ  \\
wanghu@rowan.edu}
\and
\IEEEauthorblockN{5\textsuperscript{th} Nikhil Muralidhar}
\IEEEauthorblockA{\textit{Computer Science} \\
\textit{ Stevens Institute of Technology}\\
Hoboken, NJ \\
nmurali1@stevens.edu}

}


\newcommand{\nikhilc}[1]{\textcolor{red}{\,Nikhil says: #1}}

\maketitle

\begin{abstract}

Modulation classification is a very challenging task since the signals intertwine with various ambient noises. Methods are required that can classify them without adding extra steps like denoising, which introduces computational complexity. In this study, we propose a vision transformer (ViT) based model named NMformer to predict the channel modulation images with different noise levels in wireless communication. Since ViTs are most effective for RGB images, we generated constellation diagrams from the modulated signals. The diagrams provide the information from the signals in a 2-D representation form. We trained NMformer on $106,800$ modulation images to build the base classifier and only used $3,000$ images to fine-tune for specific tasks. Our proposed model has two different kinds of prediction setups: in-distribution and out-of-distribution. Our model achieves $4.67\%$ higher accuracy than the base classifier when finetuned and tested on high signal-to-noise ratios (SNRs) in-distribution classes. Moreover, the fine-tuned low SNR task achieves a higher accuracy than the base classifier. The fine-tuned classifier becomes much more effective than the base classifier by achieving higher accuracy when predicted, even on unseen data from out-of-distribution classes. Extensive experiments show the effectiveness of NMformer for a wide range of SNRs.

\end{abstract}


\section{Introduction}
\IEEEPARstart{A}{utomatic} modulation classification (AMC) is a critical technology within wireless communication, playing a pivotal role in cognitive radio, spectrum sensing, and interference identification. Accurately recognizing modulation schemes in a communication system facilitates more efficient spectrum utilization and enhances the adaptability and efficacy of wireless networks. Achieving successful modulation classification has posed a longstanding challenge for the research community, primarily due to various factors influencing the channels, such as ambient noise. The earlier approaches to modulation classification use traditional machine learning and deep learning networks such as convolution neural networks (CNNs) \cite{peng2017modulation,8963964,10139474}. In the domain of modulation classification, Peng et al. \cite{peng2017modulation} transformed raw modulated signals into constellation diagrams, employing these diagrams as inputs for AlexNet, a CNN-based architecture. Conversely, an alternative investigation \cite{8963964} introduced MCNet, a CNN model distinguished by its utilization of multiple convolutional blocks. Notably, incorporating skip connections between these blocks effectively mitigates the issue of vanishing gradients, a common challenge in deep learning architectures. Moreover, each convolutional block is augmented with asymmetric kernels strategically crafted to capture the spatio-temporal correlations inherent within the signals under analysis.

Similarly, in a separate study detailed in \cite{10139474}, a CNN framework was devised for the AMC task. Here, the authors leveraged various signal representations, including constellation diagrams, ambiguity function (AF), and eye diagrams. Among these representations, empirical evaluations during testing revealed that constellation diagrams outperformed AF and eye diagrams regarding classification accuracy. Furthermore, a notable observation emerged from the study: combining multiple signal representations, such as constellation and eye diagrams, yielded improved classification accuracy compared to individual representations.

Attention-based models such as transformers have been popular for AMC in recent years \cite{dao2023vt, hamidi2021mcformer, cai2022signal}. Vaswani et al. pioneered transformers \cite{vaswani2017attention}, a deep learning model that excels at processing sequential data like text, speech, and DNA sequences. Transformers' main innovation is using an ``attention" mechanism, which enables the model to dynamically focus on the most relevant parts of the input while producing the output. Transformer-based architectures have garnered considerable interest in AMC due to their ability to discern robust modulation features from input data \cite{cai2022signal, kong2023transformer, qu2024enhancing}. Cai et al. \cite{cai2022signal} introduced transformers to the domain of AMC. They presented a transformer network (TRN)-based approach that surpasses conventional methods like CNNs in capturing extensive dependencies in sequential data, which is pivotal in modulation classification tasks. This makes them especially well-suited to modulation classification, where these dependencies are critical for distinguishing between different modulation schemes. The proposed method accepts preprocessed signal sequences as input and routes them to a TRN encoder. The encoder's self-attention mechanism enables it to learn long-term dependencies effectively. A multi-layer perceptron (MLP) head classifies the encoded signal representation. This method has potential advantages over traditional methods, particularly in low SNR scenarios. 

Kong et al. \cite{kong2023transformer} improved transformer-based AMC by incorporating convolutional embedding, attention bias, and attention pooling into deep neural networks. Their approach entails first training transformer encoders on unlabeled data using contrastive learning and data augmentation, followed by fine-tuning on a small labeled dataset. Their model outperforms conventional methods by incorporating convolutional elements tailored for radio signal data, especially useful in real-world scenarios with limited labeled data. To address the challenges of applying transformers to AMC tasks with limited labeled data, they use a deep neural network with convolutional embedding, attention bias, and attention pooling. In a complementary effort, Qu et al. \cite{qu2024enhancing} addressed the challenge of capturing long-range dependencies within signals by proposing a hybrid architecture combining transformers and long short-term memory (LSTM) networks. Transformers capture global correlations, whereas LSTMs are adept at handling temporal dependencies. To improve robustness against irrelevant signal characteristics, they introduce segment substitution (SS) as a new data augmentation technique. SS selectively manipulates signal segments to emphasize modulation-specific features while mitigating the influence of irrelevant factors, such as channel effects, thus augmenting the model's discriminative capacity.



Nevertheless, the most available methods \cite{peng2018modulation, 8454504} deal with classifying signals generated in deal conditions, meaning no noise is present. This approach is not feasible for practical use as signals often get mixed with noise through the medium. Moreover, few existing approaches \cite{hong2017automatic, ramjee2019fast} have a solution for fine-tuning downstream tasks with a smaller dataset. To address these issues, in this study, we developed a ViT model for AMC tasks. The main contributions of this paper are as follows:

\begin{itemize}
    \item We generated RGB constellation diagrams from modulation signals with a wide range of SNRs.
    \item We applied a ViT, named NMformer, to classify the modulation data. The proposed method uses the powerful self-attention mechanism to make decisions conditioned on global image characteristics in contrast to local image characteristics used in traditional CNN-based approaches. Our approach achieves high accuracy even for low SNR samples. 
    \item We trained a base classifier and fine-tuned it for downstream tasks. The proposed model obtains higher F1 scores for all the downstream tasks. 
\end{itemize}

The rest of this paper is organized as follows: Section II introduces the ViT for classification. Section III discusses the creation of the modulation constellation diagrams. Sections IV and V discuss the experimental setup and results. Finally, the conclusion is given in Section VI. 

\section{NMformer for Classification}

\begin{figure*}[!ht]
  \centering  
  \includegraphics[width=\linewidth]{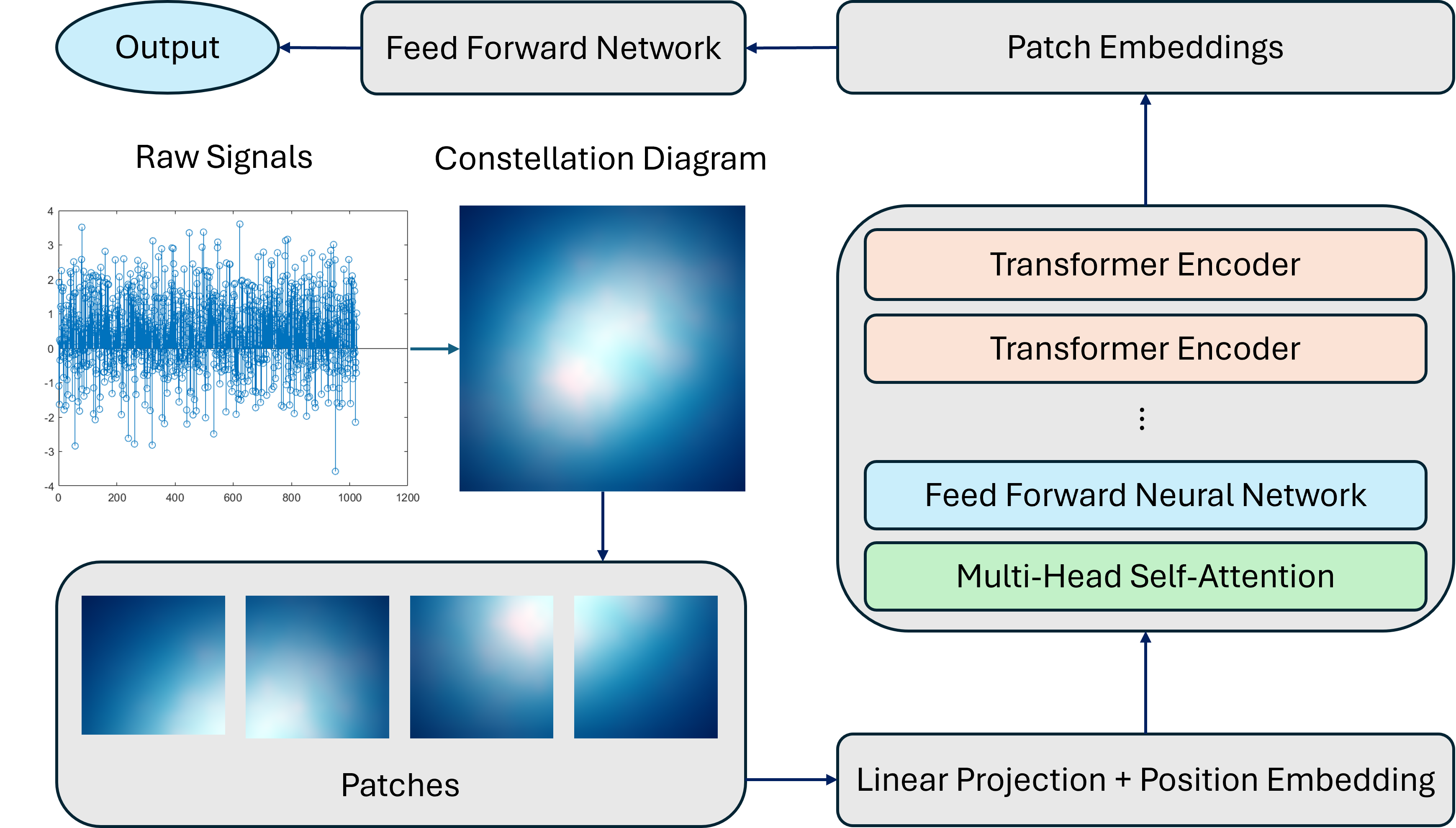}
  \caption{NMformer system sequential pipeline: starting with generating constellation diagrams from raw signals and culminating in modulation classification using a ViT architecture.}
  \label{fig:trans}
\end{figure*}

We employed a ViT for modulation classification, capitalizing on its demonstrated effectiveness in image classification tasks. The following section offers a concise overview of the ViT architecture, elucidating its pivotal role in our approach. We also depict our approach in Fig. \ref{fig:trans}. Our implementation can be found at \href{https://github.com/atik666/NMformer}{https://github.com/atik666/NMformer}.

\subsection{NMformer Architecture}

We employ a ViT architecture as the backbone of NMformer. When employing a ViT for image classification tasks in computer vision, the transformer model takes input data as two-dimensional images. The image, characterized by its dimensions of height \( H \), width \( W \), and \( C \) channels, undergoes partitioning into smaller two-dimensional patches. This partitioning yields \( N = \frac{HW}{P^2} \) patches, each having a size of \( (P, P) \) pixels.

Before feeding the data into the transformer, each image patch gets flattened into a vector \( \mathbf{x}_{p}^{n} \) of length \( P^{2}C \times 1 \), where \( n=1,\cdots, N \). Subsequently, a sequence of embedded image patches is generated by projecting these flattened patches to higher dimensions using a trainable linear projection \( E \). At the beginning of this sequence, a learnable class embedding \( x_{\text{class}} \) is introduced. This \( x_{\text{class}} \) value acts as the representation for the classification output \( y \). Furthermore, the patch embeddings are enhanced with one-dimensional positional embeddings \( E_{\text{pos}} \), which encode positional information and are trainable during the training process. The resulting sequence of embedding vectors is as follows:

\begin{equation}
    \mathbf{z}_{0}=[x_{class};x_{p}^{1}E;\cdots ;x_{p}^{N}E]+ E_{pos}
\end{equation}

For classification tasks, \( \mathbf{z}_{0} \) serves as the initial input to the transformer encoder, which consists of a sequence of \( L \) identical layers. Following this, the \( x_{\text{class}} \) value from the output of the \( L^\text{th} \) layer of the encoder is extracted and sent to a classification head. This classification head comprises a MLP that incorporates the non-linear activation function known as the Gaussian error linear unit (GELU).

\section{Modulation Data Conversion into Images}

AMC is the process of determining the modulation scheme of a wireless communication signal. AMC is critical for the efficient use of wireless communication spectrum. Traditional methods are primarily classified as computing the likelihood functions for the received or recovered signals under different modulation hypotheses and then making a classification decision. Feature-based (FB) methods involve extracting signal features from the received wireless signal and using them to classify the modulation type\cite{dobre2007survey}. Another approach to AMC is to convert the modulation data to an image format and then classify it using deep learning techniques. In AMC image transformation-based AMR approaches encode received wireless signals as amplitude and phase (A/P) components sequences.

We categorize the modulation into ten distinct formats, encompassing 4-Amplitude-Shift Keying (4ASK), 4-Pulse Amplitude Modulation (4PAM), 8-Amplitude-Shift Keying (8ASK), 16-Pulse amplitude modulation (16PAM), Continuous-Phase Frequency-Shift Keying (CPFSK), Differential Quadrature Phase Shift Keying (DQPSK), Gaussian Frequency Shift Keying (GFSK),  Gaussian Minimum Shift Keying (GMSK), On-Off Keying (OOK), and Offset Quadrature Phase Shift Keying (OQPSK).

\subsection{Constellation Diagram}

The constellation diagram, a common representation of modulated signals, depicts signal samples as points on a complex plane. While the plane is infinite, the diagram must be finite. Thus, choosing an appropriate portion of the complex plane is crucial to avoid omitting samples due to noise or clustering issues. This paper defaults to a $7 \times 7$ complex plane, assuming a typical SNR range of 0 to 10 dB, unless specified otherwise.



\subsection{Gray Image}

As pixel density increases, a constellation diagram becomes a detailed representation of signal samples, where each sample may correspond to one or more pixels. However, limited pixel density means multiple samples might be within one pixel. It's important to note that the diagram operates as a binary image, treating pixels uniformly whether they contain one sample or multiple samples.

\subsection{Enhanced Gray Image} 
   
In grayscale images, the number of samples per pixel is considered, but two limitations persist. Firstly, the exact position of each sample within a pixel isn't taken into account. Secondly, the influence of individual samples within a pixel on its neighboring pixels is ignored. To tackle these limitations, we propose an enhanced grayscale image. This image integrates the distances between sample points and pixel centroids using an exponential decay model denoted by \( B_{i,j} \). Here, \( B_{i,j} \) represents the influence of sample point \( i \) on Pixel \( j \), \( P \) denotes the sample point's power, \( d_{i,j} \) signifies the distance between sample point \( i \) and Pixel \( j \)'s centroid, and \( \alpha \) is the exponential decay rate. The impacts of all data samples on each pixel are then aggregated, yielding the intensity values for generating the enhanced grayscale image.

\subsection{Three-Channel Image}

ViTs are typically tailored for color images, comprising three RGB channels. However, the enhanced grayscale image technique operates on a single data processing channel. To align NMformer with this approach, we introduce a three-channel image configuration. This configuration generates three enhanced grayscale images from an identical set of data samples, each employing a distinct exponential decay rate.

\section{Experimental Setup}

We begin this section by outlining the parameters for generating samples in our experiments and the classifier employed. Following this, we provide the results produced by our classifier across various experimental setups.

\subsection{Parameters Selection in Data Generation}

In all our experiments, we explore ten modulation formats. The samples are generated with dimensions of $32$ pixels for both height and width, equating to a sampling frequency of $200$ kHz. The constellation's scale for depiction is fixed at $2.5$. Furthermore, we incorporate reduced path delays of $0$, $0.0004$, and $0.0008$ seconds. The number of filter taps employed is $8$.

\subsection{Parameters Selection for Classifier}

For the NMformer, we adhere to most of the hyperparameters outlined in \cite{dosovitskiy2020image} for the network architecture. The input image size is configured to $224\times 224$, representing RGB images. Each input image is partitioned into $16$ equal square patches, which, in ViT, are regarded as tokens and processed by the transformer encoder. Each patch undergoes linear projection to the embedding dimension of $768$. The ViT model comprises $12$ transformer layers. The multi-head self-attention mechanism employs $12$ attention heads, facilitating simultaneous focus on different input segments. The dimensionality of the hidden layers within the feedforward networks in the transformer blocks is fixed at $3,072$.

The number of output classes is set to $10$ for the classification task, aligning with the ten modulation formats. We employ cross-entropy loss and an Adam optimizer with a learning rate $0.00005$. A consistent batch size of $128$ is utilized across all classifiers. The base classifier undergoes training for $50$ epochs, whereas the fine-tuned classifier is trained for $100$ epochs. We implement model saving based on the criterion of achieving the lowest validation loss or the highest validation accuracy, satisfying either condition.

\subsection{Base Classifier}

For the base classifier, we consider $11$ variations of SNRs ranging from $0$ dB to $10$ dB. Within each SNR, every training class comprises at least $1,000$ images and up to $1,100$ images. Hence, the total number of training samples amounts to $106,800$. We ensure that the validation set is generated with the same distribution as the training set, which contains samples from SNRs observed during training. Specifically, we generate samples with $4$ different SNRs for validation: $0$ dB, $4$ dB, and $8$ dB, with a total of $3,000$ samples.

During testing, we evaluate the classifier on two distinct datasets: one follows the in-sample distribution, where the data originates from the same SNRs seen during training, while the other is out-of-sample distribution data, comprising novel SNRs not encountered during training.

\subsection{Fine-tuned Classifier}

To refine the performance of the base classifier model for specific tasks with limited samples, we engage in fine-tuning the learned parameters. In this process, we freeze all layers of the NMformer except for the last linear layer. During fine-tuning, each SNR is associated with $100$ training samples for each class, contrasting with the base classifier, which has a minimum of $1,000$ samples per class. Post-fine-tuning, we evaluate the model's performance not only on novel classes but also on novel SNRs for both the base and fine-tuned classifiers.

\section{Results and Discussion}

This section presents the test results derived from the base classifier trained over $100$ epochs. However, it is noteworthy that the model attains its optimal performance around the $40^{\mathrm{th}}$ epoch. Therefore, we save the model at this stage and evaluate its performance using the saved parameters. In our experiments, the out-of-distribution data typically presents a more challenging scenario for the classifier than the in-sample distribution. The outcomes are summarized in Table \ref{table:main}.

\begin{table}
\caption{Performance of the classifier on different settings.}
\centering
\begin{tblr}{
  cell{1}{1} = {c=2}{},
  cell{1}{3} = {c=2}{},
  cell{1}{5} = {c=2}{},
  cell{2}{1} = {c=2}{},
  cell{2}{3} = {c=2}{},
  cell{2}{5} = {c=2}{},
  cell{3}{1} = {c=2}{},
  cell{3}{3} = {c=2}{},
  cell{3}{5} = {c=2}{},
  cell{4}{1} = {c=2}{},
  cell{5}{1} = {r=3}{},
  vlines = {},
  hline{1-5,8} = {-}{},
  hline{6-7} = {2-6}{},
}
Model               &                          & Base          &       & Fine-tuned    &       \\
Sample              &                          & $106,800$        &       & $3,000$          &       \\
~                   &                          & Accuracy (\%) &       & Accuracy (\%) &       \\
Distribution
  type &                          & In            & Out   & In            & Out   \\
SNRs                & 0 dB, 4 dB,
  10 dB      & 71.10         & -     & -             & -     \\
                    & 0.5 dB, 1.5
  dB, 4.5 dB & -             & 67.48 & 71.60         & 70.51 \\
                    & 5.5 dB, 7.5
  dB, 9.5 dB & -             & 72.12 & 75.77         & 71.87 
\end{tblr}
\label{table:main}
\end{table}


In the first test, we evaluated the performance of the base classifier using data from the same distribution. Specifically, the test classes included SNRs of $0$ dB, $4$ dB, and $10$ dB. The resulting test accuracy achieved in this scenario was \(71.10\%\).

Subsequently, we conducted tests on two out-of-distribution datasets. The first dataset comprises SNRs of $0.5$ dB, $1.5$ dB, and $4.5$ dB, where the lower SNRs pose greater challenges for classification. In this case, the obtained test accuracy is \(67.48\%\). Conversely, the second dataset contains SNRs of $5.5$ dB, $7.5$ dB, and $9.5$ dB, presenting less challenging classification tasks due to higher SNRs. Here, we achieved a test accuracy of \(72.12\%\).

The challenges posed by different SNR distributions are reflected in the accuracy results, where the less challenging distribution yields higher accuracy than the base classifier tested on all the in-distribution samples.

In the subsequent task, we fine-tune the base classifier using only a limited number of samples for specific SNRs. For the in-distribution SNRs of $0.5$ dB, $1.5$ dB, and $4.5$ dB, we achieve an accuracy of \(71.60\%\). Surprisingly, this surpasses the accuracy of the base classifier by \(0.50\%\), despite the samples having lower SNRs. This result demonstrates that our model can achieve higher accuracy when fine-tuned with a few yet challenging samples.

Conversely, when tested with higher SNR samples of $5.5$ dB, $7.5$ dB, and $9.5$ dB, the same fine-tuned classifier achieves an accuracy of \(71.87\%\). This accuracy is higher than that of the in-distribution samples, as higher SNR samples are less challenging to classify.

Furthermore, our test on the in-distribution samples of $5.5$ dB, $7.5$ dB, and $9.5$ dB yields the highest accuracy of all, \(75.77\%\). On the other hand, the out-of-distribution samples with SNRs of $0.5$ dB, $1.5$ dB, and $4.5$ dB achieve the lowest accuracy, which is \(70.51\%\). This lower accuracy is expected since the classifier is trained on fewer noisy samples, performing better on similar samples but experiencing a drop in accuracy for noisier ones. Nevertheless, this accuracy is \(3.03\%\) higher than that of our base classifier for the same distribution, demonstrating the superiority of our fine-tuned classifier.

\begin{figure}
\centering
  \begin{minipage}{0.2\textwidth}
    \centering
    \begin{subfigure}{\linewidth}
      \includegraphics[width=\linewidth]{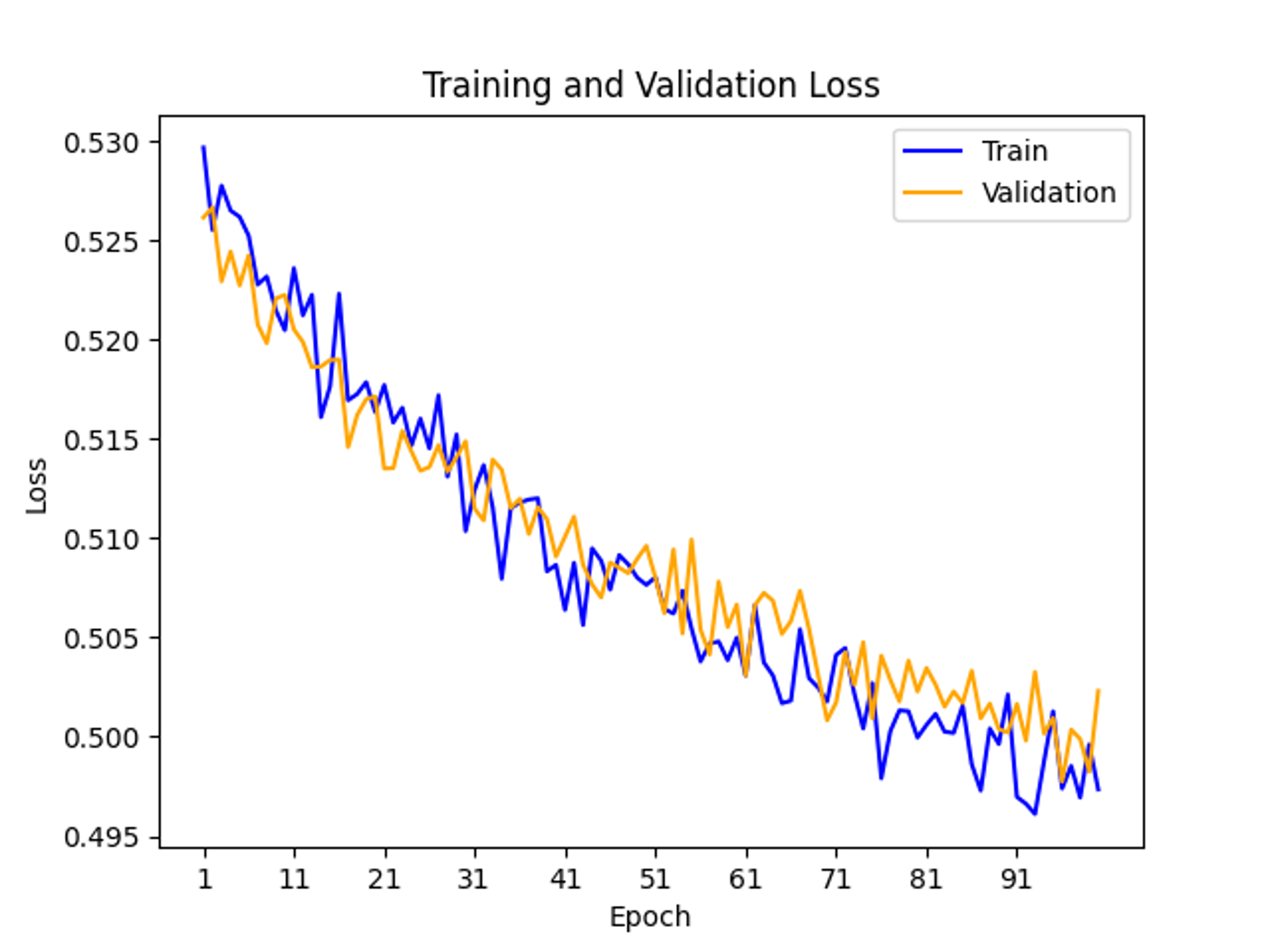}
        \caption{Loss for 0.5 dB, 1.5 dB and 4.5 dB SNRs.}
        \label{fig:sub1}
    \end{subfigure}
    \end{minipage}
    \begin{minipage}{0.2\textwidth}
    \centering
    \begin{subfigure}{\linewidth}
      \includegraphics[width=\linewidth]{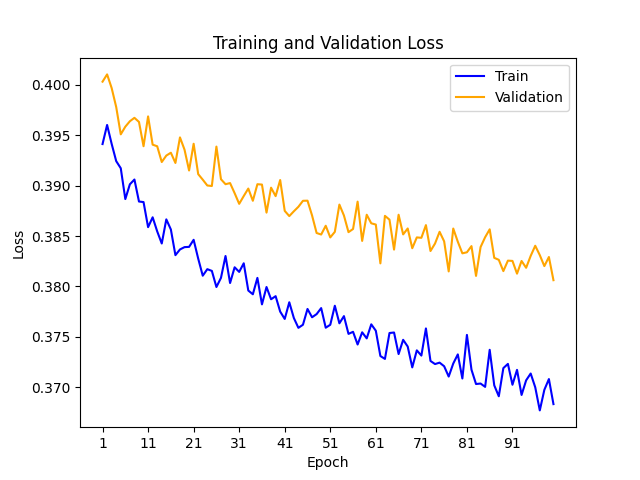}
        \caption{Loss for 5.5 dB, 7.5 dB and 9.5 dB SNRs.}
        \label{fig:sub2}
    \end{subfigure}
    \end{minipage}
    \caption{Loss convergence plot.}
    \label{fig:main}
\end{figure}

We present the model convergence plots illustrating the smooth convergence of our fine-tuned model. Figure \ref{fig:sub2} showcases the convergence behavior across different SNRs, specifically at $0.5$ dB, $1.5$ dB, and $4.5$ dB for the more challenging tasks and at $5.5$ dB, $7.5$ dB, and $9.5$ dB for the less challenging tasks.

Although the training and validation losses closely align in the first plot, they are noticeably higher compared to the latter plot. Consequently, the model achieves relatively lower accuracy at lower SNRs. However, across both plots, the model consistently converges towards the global minima as the training progresses.

In Figure \ref{fig:conf}, we present the confusion matrices for our base classifier and two fine-tuned classifiers. Notably, all samples were misclassified for classes OOK and 16PAM in the base classifier. However, in Figure \ref{fig:conf}(b), corresponding to the fine-tuned classifier, we observe a notable increase in correctly classified samples for these classes. Subsequently, in Figure \ref{fig:conf}(c), there is a significant further increase in the number of correctly classified samples, indicating the clear superiority of our fine-tuned classifier.

To dive deeper into class performance analysis, we compute the classifier's precision, recall, and F1 score, as shown in Table \ref{table:pre}. The base classifier exhibits the lowest precision, recall, and F1 score among all classifiers. The precision, recall and F1 score values show improvement for both the fine-tuned classifiers. Notably, the fine-tuned classifier with a higher SNR achieves the highest F1 score, indicating superior performance across all evaluated metrics.

{
\begin{figure}
  \centering
  \begin{minipage}{0.15\textwidth}
    \centering
    \begin{subfigure}{\linewidth}
      \includegraphics[width=\linewidth]{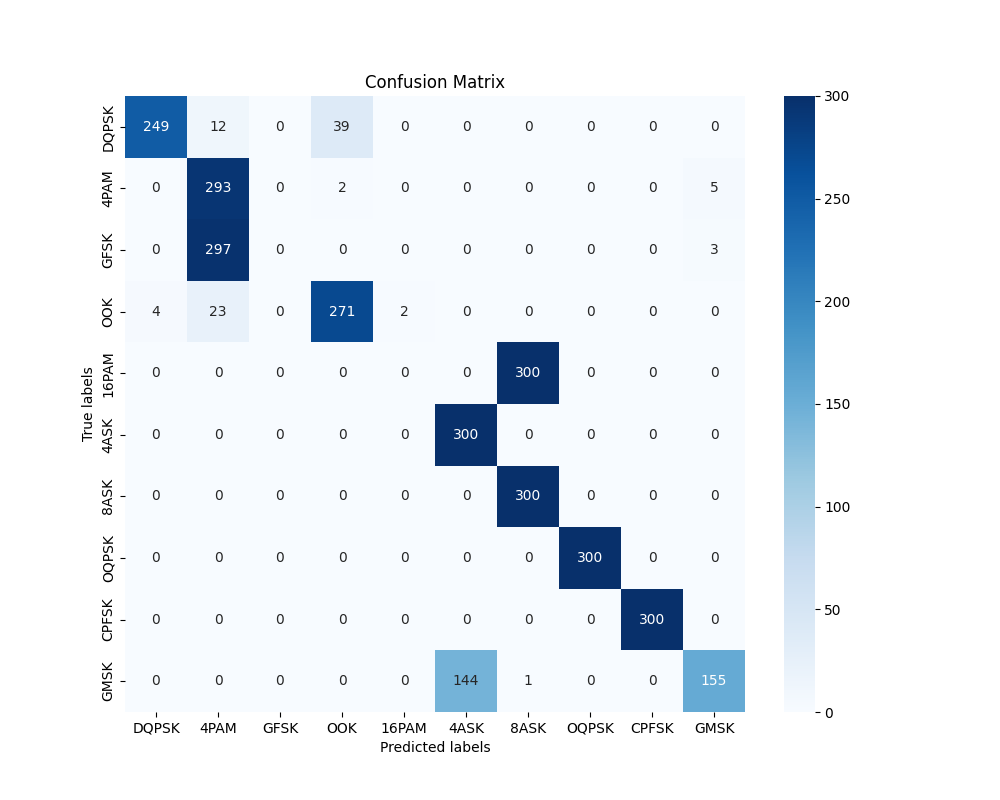}
      \caption{Base classifier.}
      \label{fig:subplot1}
    \end{subfigure}
  \end{minipage}\hspace{0.1cm}%
  \begin{minipage}{0.15\textwidth}
    \centering
    \begin{subfigure}{\linewidth}
    \vspace{0.3cm}
      \includegraphics[width=\linewidth]{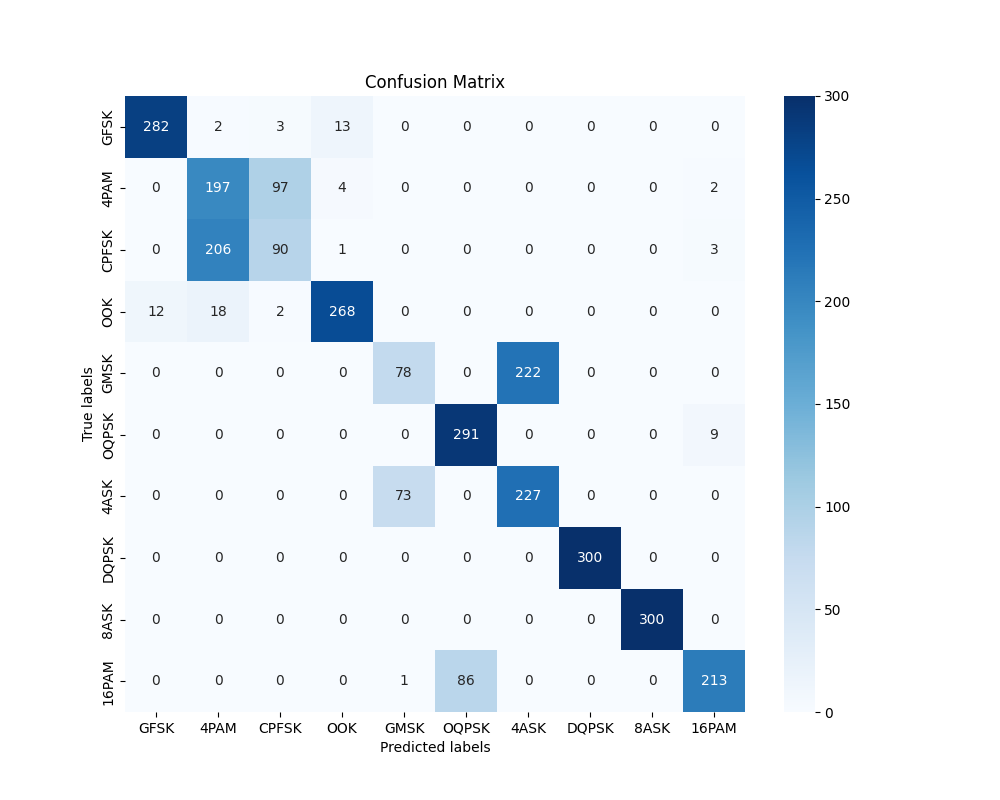}
      \caption{Fine-tuned (0.5 dB, 1.5 dB, 4.5 dB).}
      \label{fig:subplot2}
    \end{subfigure}
  \end{minipage}\hspace{0.1cm}%
  \begin{minipage}{0.15\textwidth}
    \centering
    \begin{subfigure}{\linewidth}
    \vspace{0.6cm}
      \includegraphics[width=\linewidth]{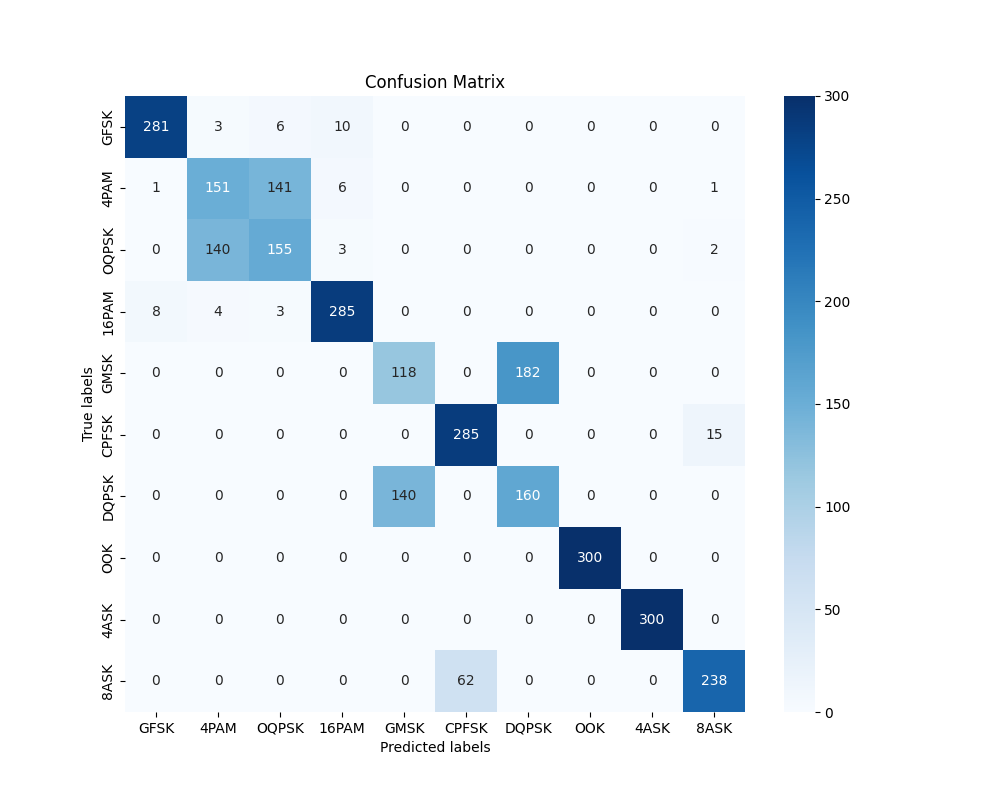}
      \caption{Fine-tuned (5.5 dB, 7.5 dB and 9.5 dB).}
      \label{fig:subplot3}
    \end{subfigure}
  \end{minipage}
  \caption{Confusion matrix for different classifiers.}
  \label{fig:conf}
\end{figure}
}
\begin{table} 
\caption{Analysis of classifier performance.}
\begin{tabular}{llll}
\hline
Classifier                             & Precision & Recall  & F1 Score \\
\hline
Base (0 dB – 10 dB)                    & 0.6170    & 0.6933 & 0.6291   \\
Fine-tuned (0.5 dB, 1.5 dB, 4.5 dB) & 0.7188    & 0.7163 & 0.7118   \\
Fine-tuned (5.5 dB, 7.5 dB, 9.5 dB) & 0.7598    & 0.7577 & 0.7572  \\
\hline
\end{tabular}
\label{table:pre}
\end{table}

\section{Conclusion}

In this study, we presented an effective model named NMformer for modulation classification in wireless communication systems leveraging a ViT model. Our proposed AMC method achieves remarkable accuracy across diverse SNRs by training on constellation diagrams of modulation signals. We demonstrate substantial improvement in classification performance by carefully fine-tuning the base classifier with limited samples tailored to specific SNRs. Our method surpasses the baseline classifier's accuracy and exhibits superior resilience to out-of-distribution data, particularly in challenging low SNR scenarios. The efficacy of our approach is further underscored by the precision, recall, and F1 score analyses, which consistently show significant improvements in classification metrics, particularly for classes traditionally prone to misclassification. Overall, our findings underscore the potentiality of NMformer for classifying noisy modulation signals.

\bibliographystyle{ieeetr}
\bibliography{egbib}

\end{document}